# Trust regulation in Social Robotics: From Violation to Repair

Matouš Jelínek

The Department of Media, Design, Education and Cognition, University of Southern Denmark, Sonderborg, Denmark, matous@sdu.dk

Kerstin Fischer

The Department of Media, Design, Education and Cognition, University of Southern Denmark, Sonderborg, Denmark, kerstin@sdu.dk

While trust in human-robot interaction is increasingly recognized as necessary for the implementation of social robots, our understanding of regulating trust in human-robot interaction is yet limited. In the current experiment, we evaluated different approaches to trust calibration in human-robot interaction. The within-subject experimental approach utilized five different strategies for trust calibration: proficiency, situation awareness, transparency, trust violation, and trust repair.

We implemented these interventions into a within-subject experiment where participants (N=24) teamed up with a social robot and played a collaborative game. The level of trust was measured after each section using the Multi-Dimensional Measure of Trust (MDMT) scale. As expected, the interventions have a significant effect on i) violating and ii) repairing the level of trust throughout the interaction. Consequently, the robot demonstrating situation awareness was perceived as significantly more benevolent than the baseline.

## 1 INTRODUCTION

In the context of human-robot interactions, trust is an essential requirement for the successful acceptance of such systems [9], as it directly influences users' willingness to engage with and rely on robotic technologies, which in turn impacts the integration of these systems into various sectors of society. As robotic systems progress into areas involving highly vulnerable populations, such as socially assistive robots employed in elderly care or healthcare settings [5], the robots must be trusted not only for their specific functionalities but also as social entities [11]. The ultimate aim is to achieve calibrated trust, a state in which the user's trust corresponds to the system's actual capabilities [19][1][4]. Miscalibrated trust levels can lead to technology rejection [7][8] and, in extreme cases, severe or even fatal accidents – e.g., [2].

In automated systems, trust has been addressed ever since more sophisticated interactions with real humans were possible, and consequently, problems with miscalibration started emerging, e.g., [15][12][2]. These discrepancies between the level of trust in the automated system and its actual capabilities have two extremes: i) over-trust and ii) under-trust.

Over-trust i) is a situation in which the trust exceeds the system's capabilities. [12] Previous research [13][16] indicates that users who collaborate with autonomous systems tend to overly rely on these systems, assigning them abilities they do not have, which may potentially lead to dangerous or even fatal situations [12].

Under-trust ii), in contrast, represents a state in which a robot's capabilities exceed the user's trust, thus causing unnecessary stress and workload for the human interaction partners and ineffective use. Appropriate trust regulation is, therefore, crucial for creating efficient human-robot interactions [19].

Even though trust in human-robot interactions has become a prominent topic of discussion within the field of human-robot interaction, only little is known about the practical implementation of trust regulation. While there are three main antecedents of trust (Human-related, Robot-related, and Environment-related) [9], this work solely evaluates possible manipulation of the robot-related antecedents, specifically potential candidates for effective trust regulation.

The Human-Robot Team (HRT) trust model [1] was used as a resource for possible trust calibration strategies. Specifically, in this study, we tested the following candidates for trust calibration: S1 proficiency, S2 situation awareness, S3 transparency, S4 trust violation, and S5 trust repair. The experiment presented in this paper will thus address the following hypotheses:

- H1: The subjective level of trust can be effectively manipulated over the course of one human-robot interaction.
- H2: The subjective level of trust can be effectively regulated, e.g., built, dampened, violated, and repaired, over the course of one human-robot interaction.

## 2 METHOD

### 2.1 Experiment design

To test our hypotheses, we created an experiment in which one human participant and one humanoid robot collaboratively played a game. Utilizing a within-subject design with five different conditions, we created a series of different behaviors the robot used throughout a single interaction. Based on the assumption that the robot must first become trustworthy to dampen or violate the participants' trust, the experiment started with strategies to build trust (S1, S2), then used a strategy to dampen trust (S3), a strategy to violate trust (S4), and finally a trust repair strategy (S5). Participants were asked to evaluate the robot after each experimental section to capture the potential change in the perceived level of trust and the effect of the specific trust calibration strategies.

The experimental session was organized around the modified version of the collaborative game Activity [20], with one human participant and one social robot interacting verbally and working as a team. In the original game, the player must perform a pantomime, describe, or draw a word or a phrase given on a card while the other team members have to guess which phrase the player is presenting. In our modification, a single team – consisting of a robot and the participant – played the game. After presenting the possible options verbally at the beginning of each round, only the robot performed the pantomime or described one of the words the human participant had to guess.

We motivated our subjects to participate in the experiment by allowing them to win a chocolate bonbon. However, because trust is only relevant if there is some vulnerability and hence some risk [19], they were warned that they could lose the chocolate if they failed to guess what their robotic partner presented correctly.

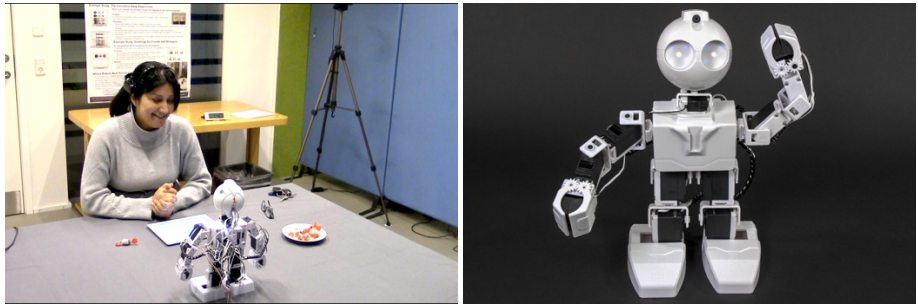

Figure 1: The setting of the experiment. Frontal image of the JD Humanoid robot used in the experiment.

*2.1.1 Candidates for trust calibration*

The gameplay was divided into five sections, each naturally accommodating one of the aforementioned trust-calibrating strategies (S1-S5). The structure of the game was the same for all participants, but with some variability to accommodate the participants' responses.

The first proficiency section (S1) focused on setting up rules and agreements [19] to demonstrate the proficiency of the robot in the interaction. The robot used utterances that demonstrate that it had the ability and power to solve the problem and could forecast and assess the situation [10].

The second section (S2) used situation awareness as a candidate for trust building, focusing on creating shared context by using strategies previously tested by [6], such as physically and verbally referring to objects in the room.

Based on the theory outlined by [19], the section focusing on transparency (S3) employed a predictive and preventative approach to trust calibration when the robot informed the participant about the limitations of its capabilities to prepare them for a potential error. Furthermore, in this experiment section, the robot revealed its weaknesses and presented how it obtains and processes information.

The trust violation approach (S4) focused on breaking the work agreement [19] and violating the previously built social norms (e.g., by being mean or suddenly playing against the human teammate). The assumption was that this would spiral into a wave of increased negativity, damaging the level of trust [19].

The final section of the gameplay (S5) focused on trust repair. An apology strategy was implemented. Initially, the robot admitted the faults and showed remorse for actions it did in the previous section. This was followed by a series of trustworthy actions, to prevent the "cheap-talk" phenomenon [17].

*2.1.2 Experimental procedure*

The experiment was conducted in the Human-Robot Interaction Laboratory at the University of Southern Denmark using the robot JD Humanoid from EZ-Robots, which is 31.8 cm tall and has 16 degrees of freedom. The robot was controlled using the Wizard-of-Oz approach [3], using the same operator for all the subjects. Our study participants were volunteers on campus in Sønderborg, who gave us consent to record the session and were informed that the robot would ask them to fill out one questionnaire after each experimental section. To test our hypothesis, we utilized the Multi-Dimensional Measure of Trust (MDMT) to elicit data regarding the subjective level of trust [18].

## 3 RESULTS

Twenty-four (24) people participated in the experiment; however, ten had to be discarded from further analysis because of technical problems with the robot, which may have influenced the course of the session and the measured level of trust. The sample analyzed included seven females and seven men; the average age was 28.33 years (sd = 9,94).

The subjective level of trust was measured after each experimental section (S1-S5) using the MDMT questionnaire [18]. A one-way repeated measures ANOVA was conducted on the 14 individuals to examine the effect that five different trust modification approaches had on the perceived level of trust. Results show that the type of strategy used led to statistically significant differences in the perceived level of trust ($F(4, 52) = 9.5447$, $p < 0.001$).

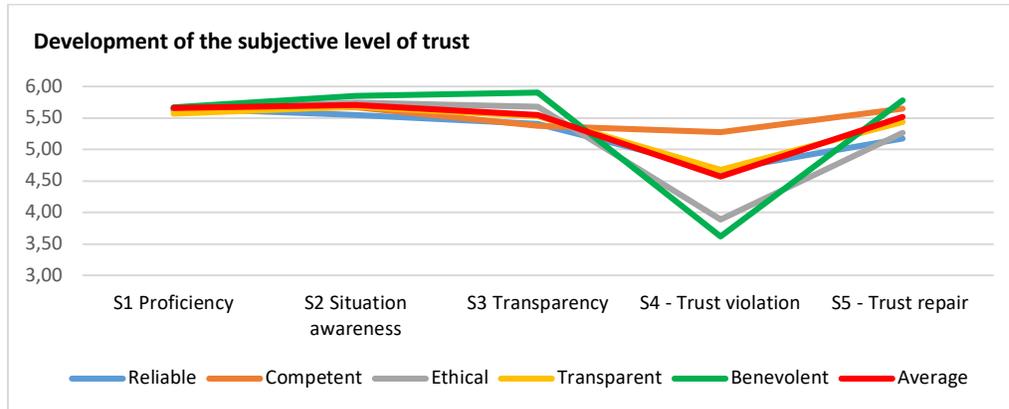

Figure 2: Results from the MDMT questionnaire, displaying the development of different components of trust in the interaction

We further analyzed the differences between sections in the following series of student t-tests. The additional analysis confirmed a significant effect of the trust violation compared with the initial stage ($p=0.005$). The subjective level of trust after the trust repair intervention (S5) was also significantly higher ($p=0.01$) compared to the measurement after trust violation (S4), indicating that our approach to violating and repairing the perceived level of trust was successful.

In addition, the different trust factors [18] were analyzed. While the performance trust remained without significant change ($p=0.07$) after the trust violation attempt, the level of moral trust dropped significantly ($p=0.003$) compared to S1. Furthermore, as expected, the robot was found to be significantly more benevolent when using strategies of shared context (S2) ($p=0.05$) compared to the proficiency stage (S1).

## 4 DISCUSSION AND FUTURE WORK

The experiment confirmed that the participants' subjective levels of trust could be successfully manipulated throughout the interaction. The robot could violate the subjective level of trust and repair it in the following experiment stage. Our main hypothesis (H1) was thus confirmed. Furthermore, we evaluated a list of candidates for future trust calibration (H2) and gathered some design recommendations for future research.

In future work, we will focus on individual trust calibration candidates. Although we identified compelling effects regarding trust violation in our experiment, from the perspective of trust calibration, future research should focus more on candidates with higher practical potential, such as transparency. Furthermore, the dynamic changes in participants' trust levels identified using a within-subject design illustrate one of the other major obstacles to the successful implementation of continuous trust regulation in human-robot interactions – the current lack of continuous online trust level measurement.

Trust regulation in Social Robotics: From Violation to Repair	SARs: TMI at CHI 2023, April 28, 2023, Hamburg, Germany# REFERENCES

[1] Anthony L. Baker, Elizabeth K. Phillips, Daniel Ullman, and Joseph R. Keebler. 2018. Toward an Understanding of Trust Repair in Human-Robot Interaction: Current Research and Future Directions. *ACM Trans. Interact. Intell. Syst.* 8, 4 (November 2018), 1–30. DOI:https://doi.org/10.1145/3181671

[2] Michael Barrett, Eivor Oborn, Wanda J. Orlikowski, and JoAnne Yates. 2012. Reconfiguring Boundary Relations: Robotic Innovations in Pharmacy Work. *Organization Science* 23, 5 (2012), 1448–1466.

[3] Christoph Bartneck. 2020. *Human-robot interaction: an introduction*. Cambridge University Press, Cambridge, United Kingdom ; New York, NY, USA.

[4] Nadine Bender, Samir El Faramawy, Johannes Kraus, and Martin Baumann. 2021. *The role of successful human-robot interaction on trust -- Findings of an experiment with an autonomous cooperative robot*.

[5] David Feil-Seifer and Maja Matarić. 2005. *Defining Socially Assistive Robotics*. DOI:https://doi.org/10.1109/ICORR.2005.1501143

[6] Kerstin Fischer, Lars Jensen, and Nadine Zitzmann. 2021. In the same boat: The influence of sharing the situational context on a speaker's (a robot's) persuasiveness. *Interaction Studies. Social Behaviour and Communication in Biological and Artificial Systems* 22, (December 2021), 488–515. DOI:https://doi.org/10.1075/is.00013.fis

[7] Mahtab Ghazizadeh, John D. Lee, and Linda Ng Boyle. 2012. Extending the Technology Acceptance Model to assess automation. *Cogn Tech Work* 14, 1 (March 2012), 39–49. DOI:https://doi.org/10.1007/s10111-011-0194-3

[8] Ella Glikson and Anita Woolley. 2020. Human trust in artificial intelligence: Review of empirical research. Academy of Management Annals (in press). *The Academy of Management Annals* (April 2020).

[9] Peter A. Hancock, Deborah R. Billings, Kristin E. Schaefer, Jessie Y. C. Chen, Ewart J. de Visser, and Raja Parasuraman. 2011. A Meta-Analysis of Factors Affecting Trust in Human-Robot Interaction. *Hum Factors* 53, 5 (October 2011), 517–527. DOI:https://doi.org/10.1177/0018720811417254

[10] Brett W. Israelsen and Nisar R. Ahmed. 2019. Dave...I can assure you ...that it's going to be all right ..." A Definition, Case for, and Survey of Algorithmic Assurances in Human-Autonomy Trust Relationships. *ACM Comput. Surv.* 51, 6 (leden 2019), 113:1-113:37. DOI:https://doi.org/10.1145/3267338

[11] Allison Langer, Ronit Feingold Polak, Oliver Mueller, Philipp Kellmeyer, and Shelly Levy-Tzedek. 2019. Trust in Socially Assistive Robots: Considerations for use in Rehabilitation. *Neuroscience & Biobehavioral Reviews* 104, (July 2019). DOI:https://doi.org/10.1016/j.neubiorev.2019.07.014

[12] John D. Lee and Katrina A. See. 2004. Trust in Automation: Designing for Appropriate Reliance. *Hum Factors* 46, 1 (March 2004), 50–80. DOI:https://doi.org/10.1518/hfes.46.1.50_30392

[13] Michael Lewis, Katia Sycara, and Phillip Walker. 2018. The Role of Trust in Human-Robot Interaction. In *Foundations of Trusted Autonomy*, Hussein A. Abbass, Jason Scholz and Darryn J. Reid (eds.). Springer International Publishing, Cham, 135–159. DOI:https://doi.org/10.1007/978-3-319-64816-3_8

[14] Roger C. Mayer, James H. Davis, and F. David Schoorman. 1995. An Integrative Model of Organizational Trust. *The Academy of Management Review* 20, 3 (1995), 709–734. DOI:https://doi.org/10.2307/258792

[15] Chang S. Nam and Joseph B. Lyons. 2020. *Trust in Human-Robot Interaction*. Academic Press.

[16] Paul Robinette, Wenchen Li, Robert Allen, Ayanna M. Howard, and Alan R. Wagner. 2016. Overtrust of robots in emergency evacuation scenarios. In *2016 11th ACM/IEEE International Conference on Human-Robot Interaction (HRI)*, IEEE, Christchurch, New Zealand, 101–108. DOI:https://doi.org/10.1109/HRI.2016.7451740

[17] Oliver Schilke, Martin Reimann, and Karen S. Cook. 2021. Trust in Social Relations. *Annual Review of Sociology* 47, 1 (2021), 239–259. DOI:https://doi.org/10.1146/annurev-soc-082120-082850

[18] Daniel Ullman and Bertram F Malle. 2020. MDMT: Multi-Dimensional Measure of Trust. (2020), 3.

[19] Ewart J. de Visser, Marieke M. M. Peeters, Malte F. Jung, Spencer Kohn, Tyler H. Shaw, Richard Pak, and Mark A. Neerincx. 2020. Towards a Theory of Longitudinal Trust Calibration in Human–Robot Teams. *Int J of Soc Robotics* 12, 2 (May 2020), 459–478. DOI:https://doi.org/10.1007/s12369-019-00596-x

[20] Activity. *BoardGameGeek*. Retrieved September 14, 2022 from https://boardgamegeek.com/boardgame/8790/activity
5